\documentclass[letter]{aa} 

\usepackage{graphicx}
\usepackage{xcolor}
\usepackage{txfonts}
\usepackage{booktabs}
\usepackage[switch]{lineno}

\begin{document}

   \title{Magnetic braking below the cataclysmic variable period gap and the observed dearth of period bouncers}


   \author{Arnab Sarkar
          \inst{1}, Antonio C. Rodriguez\inst{2}, Sivan Ginzburg\inst{3}, Lev Yungelson\inst{4}
          \and
          Christopher A. Tout\inst{1}
          }

   \institute{Institute of Astronomy, The Observatories, Madingley Road, Cambridge CB3 OHA, UK\\
              \email{as3158@cam.ac.uk}
              \and
              Department of Astronomy, California Institute of Technology, Pasadena, CA 91125, USA
\and
Racah Institute of Physics, The Hebrew University, Jerusalem 91904, Israel
\and
Institute of Astronomy of the Russian Academy of Sciences, 48 Pyatnitskaya Str.,119017 Moscow, Russia}
\titlerunning{Magnetic braking below the period gap and the period bouncer problem}
\authorrunning{Sarkar et al.}

   \date{Received xx; accepted xx}

 
  \abstract
   {Period bouncers are cataclysmic variables (CVs) that have evolved past their orbital period minimum. The strong disagreement between theory and observations of the relative fraction of period bouncers is a severe shortcoming in the understanding of CV evolution.}
   {We test the implications of the hypothesis that magnetic braking (MB), which is suggested to be an additional angular momentum loss (AML) mechanism for CVs below the period gap ($P_\mathrm{orb}\lesssim 120$ min), weakens around their period minimum.}
   {We computed the evolution of CV donors below the period gap using the MESA code, assuming that the evolution of the system is driven by AML \textcolor{black}{due to} gravitational wave radiation (GWR) and MB. We parametrised the MB strength as $\mathrm{AML_{MB}}=\kappa\mathrm{AML_{GWR}}$. \textcolor{black}{We computed two qualitatively different sets of models, one in which $\kappa$ is a constant and another in which $\kappa$ depends on stellar parameters in such a way that the value of $\kappa$ decreases as the CV approaches the period minimum ($P_\mathrm{orb}\approx80\,$ min), beyond which 
$\kappa\approx0$.}}
   {We find that \textcolor{black}{two crucial effects drive the latter set of models. (1) A decrease in $\kappa$ as CVs approach the period minimum stalls their evolution so that they spend a long time in the observed period minimum spike ($80\lesssim P_\mathrm{orb}/\,\mathrm{min}\lesssim 86$). {Here, they become difficult to distinguish from pre-bounce systems in the spike.} (2) A strong decrease in the mass-transfer rate makes them virtually undetectable as they evolve further. So, the CV stalls around the period minimum and then `disappears'. This reduces the number of detectable bouncers.} Physical processes, such as dynamo action, white dwarf magnetism, and dead zones, may cause such a weakening of MB at short orbital periods.}
   {The weakening MB formalism \textcolor{black}{provides a possible solution} to the problem of the dearth of \textcolor{black}{detectable} period bouncers in CV observational surveys.}


   \maketitle
%

\section{Introduction}
\label{sec:intro}

One of the most important challenges in our understanding of the evolution of cataclysmic variables (CVs, \citealt{2003cvs..book.....W}) is period bouncers. These are CVs that, according to the theory of CV evolution, widen their orbital separation after reaching a minimum in their orbital period, $P_\mathrm{orb}$, owing to an interplay between their mass-loss timescale, thermal timescale, and degeneracy \citep{1981ApJ...248L..27P}. However, the predicted fraction of period bouncers (70\% by \citealt{1993A&A...271..149K}, 40\% by 
\citealt{2015ApJ...809...80G}) is much greater than that inferred observationally (14\% by 
\citealt{2020MNRAS.494.3799P}, a few percent by \citealt{2023MNRAS.525.3597I}).

\cite{2020MNRAS.494.3799P} point out that current models of CV evolution (e.g. \citealt{Knigge2011}) possibly do not correctly describe their evolution.\footnote{\textcolor{black}{We note that, currently, virtually all theoretical population studies of CVs treat AML by MB,
following the law that extrapolates empirically derived time-dependence of rotational 
velocities of $\approx$10\,km/s of single stars \citep{1972ApJ...171..565S} to components of CVs with rotation velocities of $\approx$100\,km/s \citep{1981A&A...100L...7V,1983ApJ...275..713R}.}} An important ingredient governing the evolution of such CVs is magnetic braking (MB). It is well established now that there may be a mechanism of angular momentum loss (AML) operating below the period gap ($P_\mathrm{gap}$, $2\lesssim P_\mathrm{orb}\,/\mathrm{hr}\lesssim3$) in addition to AML by gravitational wave radiation (GWR). This is because the period minimum, $P_\mathrm{min}\approx 70\,$min, by a system evolved solely with $\mathrm{AML_{GWR}}$ (\citealt{2016ApJ...833...83K}) disagrees with observations, which find that $P_\mathrm{min}\approx 80\,$min \citep{Gnsicke2009}. \textcolor{black}{\cite{Knigge2011} suggested that the existence of an additional AML below the period 
gap that is stronger by a factor of 1.47  
than the $\mathrm{AML_{GWR}}$ can reproduce the  $P_\mathrm{min}$ 
of CV correctly.} 
However, there is no evidence that $\mathrm{AML}$ in short-period CVs can be simply described by a scaling factor applied to $\mathrm{AML_{GWR}}$ or that this scaling factor remains constant throughout the evolution of these CVs. \textcolor{black}{Changing the AML strength of CVs at short periods has strong implications not only on their evolution but also on their detectability. The latter strongly depends on the inferred mass-transfer rate of the system (Appendix~\ref{app:accretion}). 
The essence of our proposed solution to the dearth of observed period bouncers is that although period bouncers exist, they are either difficult to distinguish from pre-bouncers, or simply not detectable.}

In Sect.~\ref{sec:paradigm}, we describe our weakening MB paradigm and illustrate its results. In Sect.~\ref{sec:physics}, we discuss the physical processes that can weaken MB in short-period CVs. We conclude in Sect.~\ref{sec:conclusion}.

\section{The weakening magnetic braking paradigm}
Here, we describe our approach to study the implications of an MB strength that weakens around the CV period minimum.
\label{sec:paradigm}
   \begin{figure*}
   \centering
   \includegraphics[width=1\textwidth]{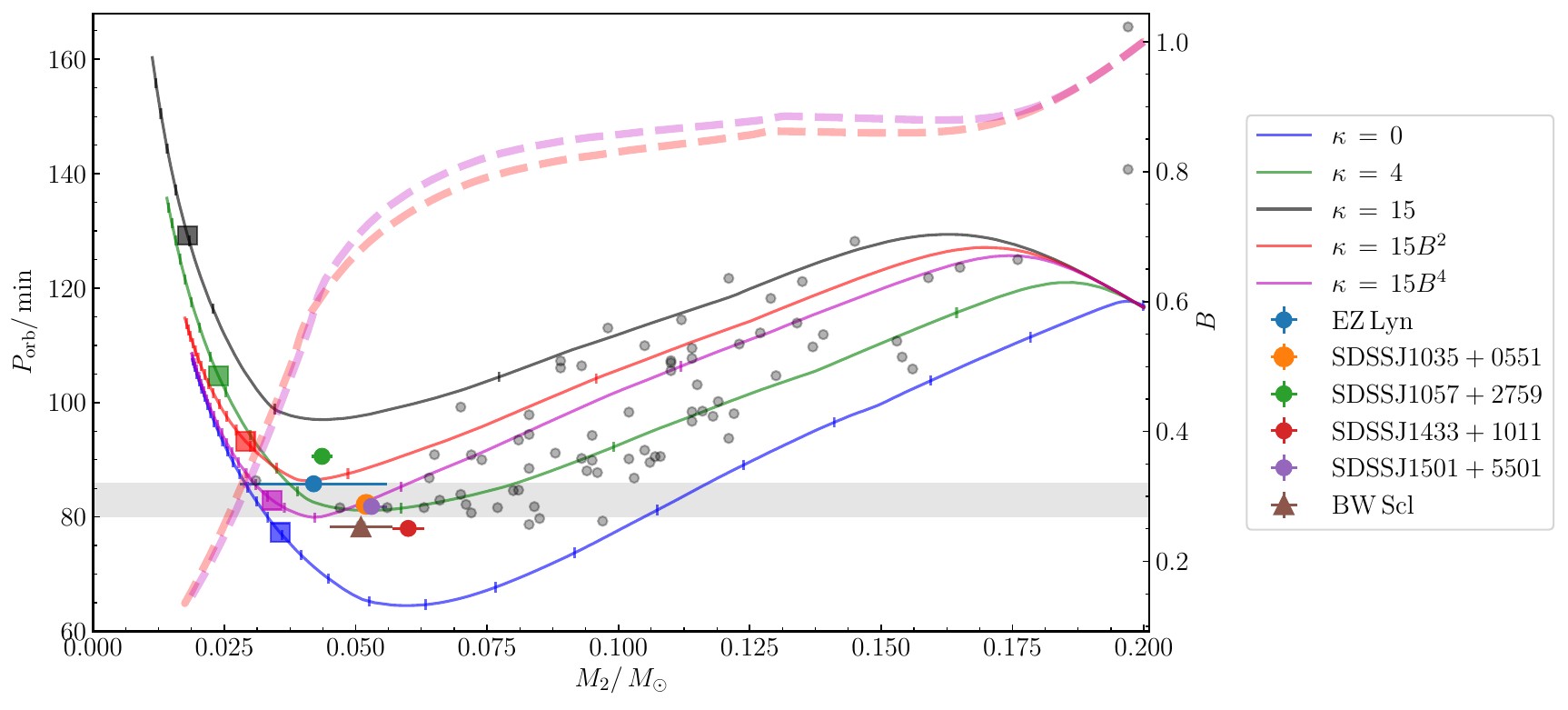}
   \caption{Evolution of CVs below the period gap. The solid lines show the tracks on the $M_2-P_\mathrm{orb}$ plane (lower x- and left y-axis). The dashed tracks in the $M_2-B$ plane (lower x- and right y-axis) show the evolution of $B$ (Eq.~(\ref{eq:B})) for the models in which $\kappa$ depends on stellar parameters. The ticks on each solid track denote timesteps of 300~Myr.  The different colors correspond to different $\kappa$s. The grey points are CVs reported by \protect\citet[their Table~1]{Knigge2006}. We also plot observed period bouncer candidates from Table~\ref{tab:catalog}. Eclipsing systems are plotted as circles, while non-eclipsing systems are plotted as triangles. The horizontal shaded region is the observed period minimum spike ($80\lesssim P_\mathrm{orb}/\,\mathrm{min}\lesssim 86$) reported by \protect\cite{Gnsicke2009}. \textcolor{black}{The squares of different colors show when the system has $\dot{M}_2 = 10^{-11} M_\odot \,\mathrm{yr^{-1}}$. During further evolution, the system is presumably undetectable (see text). } }
              \label{fig:P_M}%
    \end{figure*}
\subsection{Binary evolution calculation}
\label{sec:MESA}
We computed the evolution of CVs starting from a detached system with a fully convective donor of mass $M_{2,\,i}=0.2M_\odot$, a WD accretor of mass $M_{1,\,i}=M_\mathrm{WD,\,i}=0.8M_\odot$, and an initial period of $P_\mathrm{orb,\,i}=3.18\,$hr using version r23.05.1 of MESA \citep{Paxton2011,Paxton2013,Paxton2015,Paxton2018,Paxton2019, Jermyn2023}. The system parameters were chosen such that $P_\mathrm{orb,\,i}$ is the upper limit of $P_\mathrm{gap}$ and $M_{2,\,i}$ is the donor mass at $P_\mathrm{gap}$ reported by \cite{Knigge2011}. All our results were obtained by modifying \texttt{jdot\_multiplier} in \texttt{project\_inlist} in the MESA code, which multiplies $\mathrm{AML_{GWR}}$ by the factor \texttt{jdot\_multiplier}. We call this factor $1 + \kappa$. We defined $\kappa$ as a parametrised estimate of the strength of $\mathrm{AML_{MB}}$. This approach is similar to that of \cite{Knigge2011}. However, here $\kappa$ may also depend on stellar parameters (Sect.~\ref{sec:method}). We assume fully conservative mass transfer so that the only mechanisms of AML are GWR and MB.

We note importantly that no unique $M_\mathrm{2,gap}$ exists for all CVs, as is known from observations and theoretical computations (e.g. \citealt{Knigge2011} and \citealt{2022MNRAS.513.4169S}). This is because for un-evolved CV donors, $M_\mathrm{2,gap}$ and the lower end of the period gap, at which mass transfer resumes ($P_\mathrm{gap,-}$), depend on $M_1$ and the MB strength above the period gap. If we assume that the initial strength of MB below the period gap depends on stellar parameters, $M_\mathrm{2,gap}$ and $P_\mathrm{gap,-}$ set the initial $\kappa$ of our systems. In the next section, we choose how $\kappa$ behaves as the CV evolves.

\subsection{The method}
\label{sec:method}

We considered two qualitatively different sets of models, one in which $\kappa$ is a constant throughout and another in which $\kappa$ varies with stellar parameters. In the first set, we computed models evolved with $\kappa=0,\,4,\,\mathrm{and}\,15$ (Fig.~\ref{fig:P_M}). The latter two $\kappa$s are purely ad hoc and were chosen to aid understanding of an MB that depends on stellar parameters, described later. The system with $\kappa=0$ evolves solely with GWR and represents the minimum $P_\mathrm{orb}$ of a given $M_2$. Systems such as polars (AM Her systems, \citealt{1994MNRAS.268...61L}), in which there is no MB, follow this track. In zero-age CVs with some degree of MB, $\kappa>0$ initially. We plot in Fig.~\ref{fig:P_M} CVs with donor masses and periods, estimated from superhump periods, 
assuming $M_1=0.75M_\odot$ by \citet[their Table~1]{Knigge2006}.\footnote{{These are pre-bounce CVs. The $M_2$ for post-bounce CVs was calculated differently (Sect.~\ref{subs:results})}.} There is quite a bit of scatter among these points and they do not seem to converge on a unique evolutionary track. 
This illustrates that varying strengths of MB likely operate below the period gap.\footnote{{We note that the scatter of the systems in the catalogue of \cite{Knigge2006} may be the result of other evolutionary processes which can alter the thermal timescale of the donor.}} However, studying the significance of this effect is beyond the scope of this Letter. The track with $\kappa=15$ matches with the systems with the biggest $P_\mathrm{orb}$ for a given $M_2$ in the catalogue of \cite{Knigge2006}. So, hereinafter, we assume that the tracks with $\kappa=15$ and $\kappa=0$ exhibit, respectively, the upper and lower limits of $P_\mathrm{orb}$ for a given $M_2$.

The other set of tracks, in which $\kappa$ varies with stellar parameters, illustrates the behaviour of the system when the strength of MB depends on stellar structure and {changes as the donor evolves}. We used the result of the strong- and weak-field dynamo for fully convective low-mass stars proposed by \cite{2011MNRAS.418L.133M} to model such an MB strength. They argued, based on spectropolarimetric observations by \cite{2010MNRAS.407.2269M}, that two different magnetic field profiles exist in isolated fully convective stars with similar rotation rates and masses. The first is a strong and steady axial dipole field and the second is a weak, multi-polar non-axisymmetric field that is changing rapidly. Because donors in short-period CVs are fully convective, it is possible that a strong-field dynamo also operates in such CV donors in which it drives MB. So, we used the formula for the magnetic field given by \citet[their Eq.~(2)]{2011MNRAS.418L.133M} to define $\kappa$ (for details on how they derive their magnetic field expression, we urge the reader to refer to their Sect.~4.2). Other physical mechanisms that may lead to a stellar-dependent $\kappa$ are discussed in Sect.~\ref{sec:physics}.

We defined a dimensionless quantity, $B$, as a proxy for the magnetic field as\footnote{{There is an additional term $(\eta_\odot/\eta_\mathrm{ref})^{1/2}$ in the expression of the magnetic field in \cite{2011MNRAS.418L.133M}. Here $\eta_\mathrm{ref}\equiv10^{11}\,\mathrm{cm^2\,s^{-1}}$ is the magnetic diffusivity and $\eta_\odot$ is the reference magnetic diffusivity. Studying how this term varies for our CVs is beyond the scope of this work. So we set $(\eta_\odot/\eta_\mathrm{ref})=1$. }}
\begin{equation}
    B = \frac{6\,\mathrm{kG}}{19.5\,\mathrm{kG}} \left(\frac{M_2}{M_\odot}\right)^{1/2} \left(\frac{R_2}{R_\odot}\right)^{-1} \left(\frac{L_2}{L_\odot}\right)^{1/6} \left(\frac{P_\mathrm{orb}}{\mathrm{d}}\right)^{-1/2},
    \label{eq:B}
\end{equation}
where $R_2$ and $L_2$ are the radius and the luminosity of the donor, respectively. We computed these using MESA. {The last term in Eq.~(2) of \cite{2011MNRAS.418L.133M} is $({P_\mathrm{spin}}/\,{\mathrm{d}})^{-1/2}$, where $P_\mathrm{spin}$ is the spin period of the M-dwarf. This becomes $({P_\mathrm{orb}}/\,{\mathrm{d}})^{-1/2}$ in our Eq.~(\ref{eq:B}) because of tidal locking.} The denominator 19.5 kG is the dipolar field at the time of the commencement of Roche lobe overflow (RLOF). This ensures that $B<~1$ throughout the evolution. We plot two tracks in which $\kappa=~15\,B^2\:\mathrm{and}\: \kappa=15\,B^4$. The exponents are ad hoc but highlight the varying degrees of the dependence of MB strength on the magnetic field, and hence the stellar structure. They also lead to the system attaining $P_\mathrm{min}$ at 86 and 80 min, respectively (Fig.~\ref{fig:P_M}), which are the upper and lower limits of the observed period minimum spike reported by \cite{Gnsicke2009}. {The behaviour of $B$ can be understood as follows. Because of RLOF and the fact that the donors are close to thermal equilibrium, $R_2$, $L_2$, and $P_\mathrm{orb}$ are functions of $M_2$, and so $B\equiv B(M_2)$ \textcolor{black}{and} $P_\mathrm{orb}\propto R_2^{3/2}M_2^{-1/2}$. For our donors, $L_2\propto M_2^\beta$, in which $2\lesssim\beta\lesssim4$ \textcolor{black}{depending on the mass-transfer rate}. If we define $R_2\propto M_2^\alpha$, we get $B\propto M_2^{3/4 + \beta/6 -7\alpha/4}$. We have $\alpha>0$ pre-bounce and $\alpha\lesssim0$ post-bounce. Choosing $\beta=3$, $\alpha=0.6$ pre-bounce and $\alpha=0.3$ post bounce (similar to eq.~(16) of \citealt{Knigge2011}), we get $B\propto M_2^{0.2}$ pre-bounce and $B\propto M_2^{0.725}$ post-bounce.  So, post-bounce $B$ decreases strongly because of a change in the $M_2-R_2$ relation of the donor.}  The evolution of $B$ is shown in Fig.~\ref{fig:P_M}.

\subsection{Results}
\label{subs:results}
We follow the evolution of the models with $\kappa=15B^2$ and $\kappa=15B^4$ in the $M_2-P_\mathrm{orb}$ plane. At $M_2\approx 0.2M_\odot$, these systems are driven by strong MB so they follow the track with $\kappa=15$. However, $B$ starts decreasing {gradually at $M_2\approx0.125M_\odot$ and substantially when $M_2\lesssim0.05M_\odot$}. {This leads to the weakening of the MB strength. We note, importantly, that for all our models, the absolute value of AML decreases as the CV evolves (see Appendix~\ref{app:MB}). So, the `weakening' of MB is the additional weakening of the MB strength caused by $B$ (Fig.~\ref{fig:jdot}).} The weakening of MB is such that the donor star always adjusts to it on its thermal timescale. The extent of the weakening depends on the exponent of $B$. Close to their respective $P_\mathrm{min}$, MB becomes negligible. This can be understood with Eq.~(\ref{eq:B}) — further evolution decreases $M_2$ and increases $R_2$ and, as a consequence, $P_\mathrm{orb}$. These systems, now only driven by GWR, evolve further to converge to the $\kappa=0$ track. This causes their evolution timescale to drastically increase around and beyond their $P_\mathrm{min}$. Owing to their long evolutionary timescales, these systems stall in the period minimum spike and spend a lot of time there compared to systems evolved with a constant $\kappa$. 
Because the systems are clustered around the period minimum spike, here it is very difficult to distinguish between pre-bounce and post-bounce systems observationally \citep{2018MNRAS.481.2523P}. We highlight that the weakening MB models also reproduce the period minimum reported by \cite{Knigge2011} but that the $M_2$ at which $P_\mathrm{min}$ is attained is much smaller than the 0.069$M_\odot$ reported by \cite{Knigge2011}. So, if MB weakens in near-$P_\mathrm{min}$ CVs, our models suggest that most of the period bouncer candidates in Fig.~\ref{fig:P_M} are pre-bounce CVs.

\textcolor{black}{At this stage the problem is far from over. Any MB strength below the period gap will only accelerate the evolution of short-period CVs towards their period minimum and drive more CVs to form period bouncers. This will lead to more period bouncers than are predicted solely using GWR (e.g. \citealt{1993A&A...271..149K}, \citealt{2015ApJ...809...80G}), and thus exacerbate the classical problem of the dearth of observed period bouncers. 
For a detectable period bouncer we not only need $P_\mathrm{orb}\geq P_\mathrm{min}$ and $M_2\leq M_2(P_\mathrm{min})$, but also $\dot{M}_2 > \dot{M}_{2,\,\mathrm{detect}}$, where $\dot{M}_{2,\,\mathrm{detect}}$ is the detection threshold in the mass-transfer rate, $\dot{M}_2$.}

All of our candidate bouncer CVs (Table.~\ref{tab:catalog}) 
have $\Dot{M}_2$ (estimated by Eq.~(\ref{eq:townsley}) using WD properties) about a few times $10^{-11} M_\odot\,\mathrm{yr}^{-1}$ (also see \citealt{2022pala}). So, we assume an optimistic detection threshold of $\Dot{M}_2 = 10^{-11} M_\odot\,\mathrm{yr}^{-1}$ such that any system below this limit is undetectable. \textcolor{black}{The impact of such observational selection effects have been explored in the past (e.g. \citealt{2007MNRAS.374.1495P}, \citealt{2023MNRAS.524.4867I}).} This limit is likely to change with emerging data from optical and X-ray surveys, such as SDSS-V \citep{2017sdssv} and SRG/eROSITA \citep{2021erosita, 2021sunyaev}, respectively. The former has already led to the discovery of new period bouncer candidates, which are optically fainter than much of the population \citep{2023MNRAS.525.3597I}. The latter is five to 15 times deeper than the last all-sky X-ray survey, potentially revealing systems with lower accretion rates; for instance, the bouncer candidate reported by \cite{2024arXiv240104178G}. 
\textcolor{black}{In Appendix~\ref{app:accretion}, we discuss the effects of using $\dot{M}_2$ derived from X-ray luminosity and its limitations. Later in this section we also discuss the effect of changing $\dot{M}_{2,\,\mathrm{detect}}$.}

   \begin{figure}
   \includegraphics[width=0.4\textwidth]{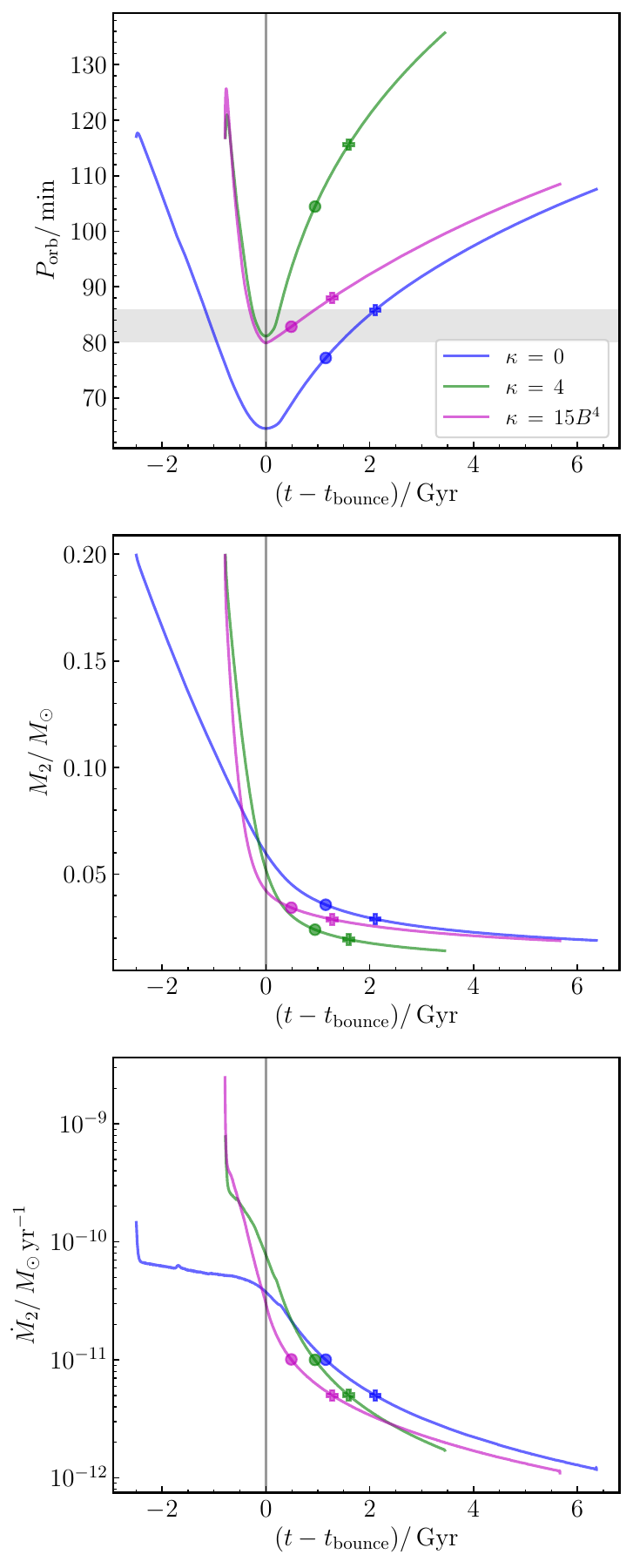}
   \caption{Time evolution of $P_\mathrm{orb}$, $M_2$, and $\dot{M}_2$ for three choices of $\kappa$. The vertical black line in each subplot is the time at which the system attains its period minimum ($t_\mathrm{bounce}$). The circles and pluses in each track mark detection thresholds of $\dot{M}_2 = 10^{-11}M_\odot\,\mathrm{yr^{-1}}$ and $ 5\times10^{-12}M_\odot\,\mathrm{yr^{-1}}$ (see text). The shaded region in the top subplot is the observed period minimum spike.}
              \label{fig:otherobs}%
    \end{figure}

\textcolor{black}{A complete picture of how the weakening of MB can reduce the number of detectable period bouncers is shown in Fig.~\ref{fig:otherobs}. We chose $\Dot{M}_{2,\,\mathrm{detect}}=10^{-11} M_\odot\,\mathrm{yr}^{-1}$ (circles in each track), beyond which the system becomes undetectable. Although $\dot{M}_{2,\,\mathrm{detect}}$ is model-independent, how a variable MB strength drives a system to reach this $\dot{M}_{2,\,\mathrm{detect}}$ is model-dependent. In the $t-\dot{M}_2$ plot, it can be seen that the $\dot{M}_2$ of the $k=15B^4$ model attains $\dot{M}_{2,\,\mathrm{detect}}$ earlier than that of the other two models. The time evolution of $P_\mathrm{orb}$ and $M_2$ is affected because of this. In the $t-P_\mathrm{orb}$ plot, the $\kappa=4$ model bounces at 80~min and becomes undetectable at 105~min at $t=0.94$~Gyr. The $\kappa=0$ model bounces at 65~min and becomes undetectable at 78 min at $t=1.15$~Gyr. The $k=15B^4$ model bounces at 80~min and becomes undetectable at 83~min at $t=0.48$~Gyr. So the weakening MB system becomes undetectable earlier and without much change in $P_\mathrm{orb}$ post-bounce (there is only about a 3~min difference between bounce and non-detection). The $\kappa=4$ and $\kappa=0$ models have a large difference between the $P_\mathrm{orb}$ in which they bounce and in which they become undetectable. The same is true in the $t-M_2$ plot. In the model with weakening MB, there is little change in its $M_2$ between bounce and non-detection, while the change is more significant for the $\kappa=0$ and $\kappa=4$ models. 
The results do not change qualitatively  if
we take  $\dot{M}_{2,\,\mathrm{detect}}=5\times10^{-12}M_\odot\,\mathrm{yr^{-1}}$ (pluses in each track), the limit we expect from SRG/eROSITA \citep{2024arXiv240104178G}. In other words, we conclude that the weakening of MB greatly slows down the evolution of the system, with lesser change in $M_2$ and $P_\mathrm{orb}$, from the time it bounces to the time it becomes undetectable. \textcolor{black}{However, we note that all our tracks asymptotically tend to $\dot{M}_2=10^{-12}M_\odot\,\mathrm{yr^{-1}}$. So, our idea makes a strong testable prediction for finding many low-luminosity bouncers aggregated at $\dot{M}_2\approx10^{-12}M_\odot\,\mathrm{yr^{-1}}$ in the upcoming surveys (e.g. \citealt{2024arXiv240104178G}). A dearth of CVs at these $\dot{M}_2$ will easily falsify this theory, although finding such low $\dot{M}_2$ might be difficult (see Appendix~\ref{app:accretion}). } The evolution of the models in the $\dot{M}_2-P_\mathrm{orb}$ plane is shown in Fig.~\ref{fig:p_mdot} and discussed in detail in Appendix~\ref{app:accretion}.}

For $\Dot{M}_{2,\,\mathrm{detect}}=10^{-11} M_\odot\,\mathrm{yr}^{-1}$, our weakening MB model ($\kappa=15 B^4$) predicts a reduction in the time spent by a system as a detectable bouncer by a factor of 2.35 compared to the GWR model. Simply reducing the fraction of bouncers predicted by solely using GWR in \citet[about 70\%]{1993A&A...271..149K} and \citet[about 40\%]{2015ApJ...809...80G} by 2.35, we get equivalent fractions of detectable period bouncers of about 30\% and 17\%, respectively. We note that the estimates of \cite{1993A&A...271..149K} and \cite{2015ApJ...809...80G} are the fraction of all period bouncers. The detectable ones are a small subset of it. Additionally, a further reduction to match the observationally inferred estimates (e.g. \citealt{2023MNRAS.525.3597I}) is possible because we show that detectable bouncers populate the period minimum spike that is also populated by pre-bounce CVs. Here, they are difficult to distinguish observationally \citep{2018MNRAS.481.2523P}. 

We note that although \cite{2023MNRAS.524.4867I, 2023MNRAS.525.3597I} show that bouncers make up only a few percent of the total CV population, this population also consists of magnetic CVs, which follow a different evolution than non-magnetic CVs \citep{1994MNRAS.268...61L}. If we assume that bouncers remain non-magnetic throughout (although see Sect.~\ref{sec:WD}), a better estimate of the fraction of bouncers amongst all non-magnetic short-period CVs can be obtained from the top panel of fig.~33 of \cite{2023MNRAS.524.4867I}. Here, we designate SU~UMa systems as pre-bounce CVs and assume that most WZ~Sge are bouncers. \textcolor{black}{These assumptions, although crude, allow us to approximate that} period bouncers make up roughly 30\% of all non-magnetic CVs below the period gap in the SDSS~I to IV catalogue. 

\begin{figure}
   \includegraphics[width=0.4\textwidth]{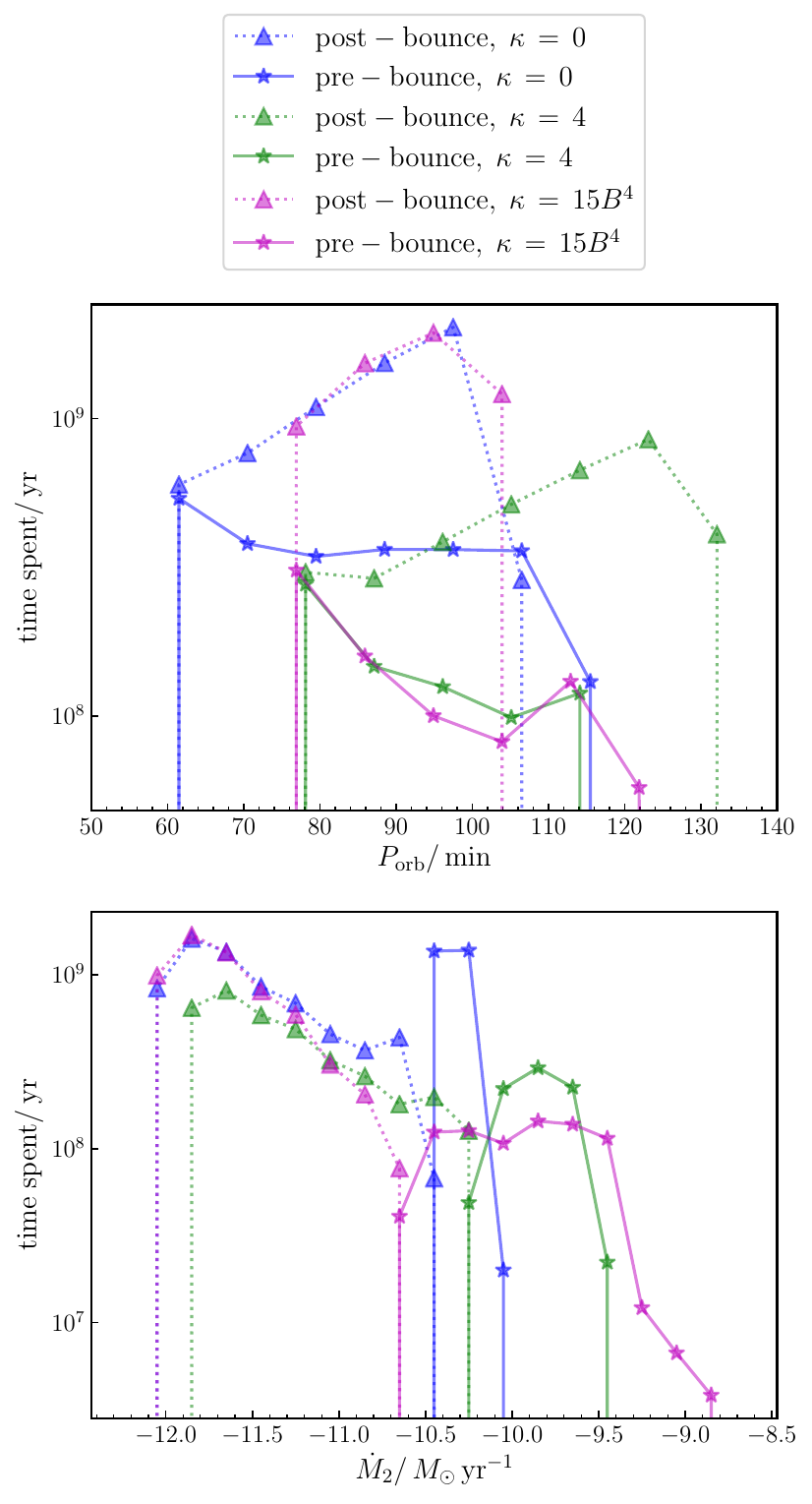}
   \caption{Distributions of the evolution time spent at each increment of $P_\mathrm{orb}$ and  $\dot{M}_2$ for three choices of $\kappa$. The stars denote pre-bouncers ($M_2>M_2(P_\mathrm{orb}=P_\mathrm{min})$) and triangles denote post-bouncers. Top panel: Systems evolving from right to left as pre-bouncers (star symbols) and from left to right (triangles) post-bounce. Bottom panel: Systems evolving from right to left.}
              \label{fig:hist}%
    \end{figure}

We perform a direct comparison of our results with that of \cite{2023MNRAS.524.4867I} using the distribution of the evolution time spent by our systems as a function of $P_\mathrm{orb}$ and $\dot{M}_2$ in Fig.~\ref{fig:hist}. Because we make no claims on the nature of MB above the period gap, we assume that all our models follow identical evolution till the lower end of the period gap. So, the time spent by a system at a given $P_\mathrm{orb}$ and $\dot{M}_2$ interval below the gap is proportional to the number of systems in that interval. As was expected, it is seen that the intrinsic distribution of short-period non-magnetic CVs is dominated by bouncers. The intrinsic fraction of bouncers is about 74\% for $\kappa=0$,  82\% for $\kappa=4$, and 87\% for $\kappa=15B^4$. This was also expected, because in the $\kappa=15B^4$ model we introduce a substantial MB at the beginning, which drives more CVs towards becoming bouncers. 

Because the $\kappa=4$ and $\kappa=15B^4$ models have similar $P_\mathrm{min}$, we analysed how their features may compare with observations. We calculated the number of bouncers that would have to become invisible below a certain $\dot{M}_2$ cut-off \footnote{These need not necessarily be our previous $\dot{M}_{2,\mathrm{detect}}$.} for our fraction of visible period bouncers to corroborate the 30\% that we obtained from fig.~33 of \cite{2023MNRAS.524.4867I}. For $\kappa=15B^4$, the limit is $\dot{M}_2 \approx 1.2\times10^{-11} M_\odot\,\mathrm{yr^{-1}}$, so that about 94\% of the intrinsic bouncers become invisible. For $\kappa=4$, the cut-off is $\dot{M}_2 \approx 3.2\times10^{-11}M_\odot\,\mathrm{yr^{-1}}$, so that about 90\% of the intrinsic bouncers become invisible. The factor of 2.7 between the $\Dot{M}_2$ cut-off in the two models may not seem dramatic, so its significance needs to be emphasised. For an $\dot{M}_2$ cut-off of $\approx 1.2\times10^{-11} M_\odot\,\mathrm{yr^{-1}}$, the fraction of bouncers in the $\kappa=4$ model increases to about 50\%, meaning that 20\% of bouncers exist for $1.2\times10^{-11}\lesssim\dot{M}_2/\,M_\odot\,\mathrm{yr^{-1}}\lesssim 3.2\times10^{-11}$.  
In other words, for any detection cut-off in $\dot{M}_2$, our weakening MB models will be more affected than the constant MB ones and corroborate observations better. 

We note that there is no reason to suggest that our analyses with an ad hoc power of $B$ in an expression of MB derived from the prescription of \cite{2011MNRAS.418L.133M} would provide a robust comparison with the detailed observational work of \cite{2023MNRAS.524.4867I,2023MNRAS.525.3597I}. In this work, we just show that for a given observational cut-off, the number of detectable bouncers reduces strongly if MB weakens post-period minimum. The strength of this reduction, in turn, depends on how strongly MB weakens. An extreme case of this is if post-minimum, MB weakens on a dynamical timescale. In such a case, the CV detaches, as is suggested by \cite{2023MNRAS.524.4867I}, and remains so for about a Gyr (see Fig.~\ref{fig:otherobs}) till GWR resumes RLOF. 

\textcolor{black}{It is important to discuss our results in the context of certain population synthesis studies; for instance, \cite{2007MNRAS.374.1495P} and \cite{2020MNRAS.491.5717B}. They find, using MB from \cite{1981A&A...100L...7V} and \cite{1983ApJ...275..713R} above the period gap, respectively, and no MB below the gap, that the intrinsic CV population cannot contain as large a fraction of short-period systems, specifically period bouncers, as is predicted by theory and that selection effects (such as our $\dot{M}_{2,\,\mathrm{detect}}$ cut-off) cannot reconcile the predictions of CV evolution theory with observations. At first glance, these results seem to defeat the main idea in this paper, in which our claim that there is an MB strength about 15 times as strong as $\mathrm{AML_{GWR}}$ below the period gap exacerbates this apparent discrepancy. However, we note that the intrinsic populations constructed by \cite{2007MNRAS.374.1495P} and \cite{2020MNRAS.491.5717B} entirely depend on their assumed MB strength above the period gap. There is evidence to suggest that a weaker MB possibly operates above the gap \citep{Knigge2011}. A weaker MB above the period gap lowers the birthrate of all CVs, which agrees with \cite{2020MNRAS.494.3799P}, who find a lower space density of CVs. Fig.~2 of \cite{Knigge2011} illustrates that the AML strengths via MB in \cite{1981A&A...100L...7V} and \cite{1983ApJ...275..713R} are about two orders of magnitude stronger than GWR above the period gap. So, no matter what strength of MB we choose below the gap, too many CVs have already been driven below the period gap by the strong MB above the gap and so a discrepancy between theory and observations is bound to arise.  What causes the discrepancy between theoretical predictions and observations is not what one chooses as MB below the period gap but what one chooses as MB above the period gap. In other words, a modest initial MB strength of about $15\mathrm{AML_{GWR}}$ below the period gap cannot undo the accumulation of CVs already dumped there by a strong MB above the gap. Because we make no claims about MB above the gap in this work, there is no reason for our idea to corroborate studies in which the theoretical CV population is entirely shaped by uncertain estimates of a strong MB above the period gap. }





\section{Physical processes driving the weakening of magnetic braking}
\label{sec:physics}
{We highlight a few physical processes that may cause the weakening of MB in short-period CVs. We note that this list is not exhaustive and that there can be additional mechanisms driving such a weakening.}
\subsection{Dynamo action in cool stars}
\label{sec:dynamo}
In Sect.~\ref{sec:method}, we showed that if the strong-field dynamo proposed by \cite{2011MNRAS.418L.133M} operates in short-period CV donors, Eq.~(\ref{eq:B}) causes $B$ to reduce significantly for $M_2\lesssim0.07M_\odot$. There is observational evidence to suggest that stars with $T_\mathrm{eff}\lesssim2200\,$K such as L-dwarfs have significantly lower chromospheric activity compared to M-dwarfs despite being rapid rotators \citep{2003ApJ...583..451M}. This means that the magnetic field strength drops from fully convective M-dwarfs to brown dwarfs. {In Sect.~\ref{sec:method}, we showed that this drop may be due to the change in the mass-radius relation of the star}. The results of the $\alpha^2$ dynamo model proposed by \cite{Chabrier2006} have also shown, similarly to \cite{2011MNRAS.418L.133M}, that there is a transition in the magnetic field structure from a steady, large-scale field in late M-dwarfs to a toroidal, oscillatory one in brown dwarfs. In addition, the conductivity of the atmosphere of cool objects such as brown dwarfs decreases greatly, thereby hampering the formation of a hot corona that would drive stellar winds. The combined effect of weaker stellar winds and reduced magnetic field strength drives a weaker MB in brown dwarfs \citep{2003ApJ...583..451M, Chabrier2006}. In other words, if such a dynamo operates in short-period CV donors, MB reduces significantly as the donor enters the brown dwarf regime ($M_2\lesssim0.07M_\odot$). 

\subsection{White dwarf magnetism}
\label{sec:WD}
{\cite{2017ApJ...836L..28I} suggest that cool WDs generate strong magnetic fields by a crystallization-driven dynamo. \cite{2021NatAs...5..648S} show that magnetic CVs can be explained by the rapid rotation and crystallization of the WD accretors, which can generate fields of several MG \citep{Ginzburg2022}. \cite{2023A&A...679L...8S} have recently proposed that such fields are generated in the accretor of short-period CVs post-period minimum. This field connects with that of the donor star, resulting in the detachment of period bouncers for several gigayears. They argue that this can lead to a reduction of about 60\% in the number of observed bouncers.}


We {illustrate a variation in their analysis in which} the CV may remain semi-detached. \cite{2023A&A...679L...8S} assume that {the diffusion timescale of the magnetic field to the WD surface is 100~Myr (Fig.~3 of \citealt{Ginzburg2022})}. However, recently \cite{2024MNRAS.528.3153B} showed that the magnetic field on the WD surface gradually emerges on a gigayear timescale (their bottom right subplot in Fig.~1). {By consistently taking into account phase separation,  they find that the magnetic diffusion time is about 1 Gyr at the time of breakout and shortly afterwards (this also depends on the WD mass)}.The donor has a thermal timescale of around a few gigayears, depending on the mass-transfer rate. {Since the thermal timescale of the donor is comparable to the diffusion timescale of the WD magnetic field,} there is a possibility that the donor adjusts to the reduction in MB because of magnetic reconnection post-period minimum, while continually filling its Roche lobe. In such a case, the evolution will be similar to that presented in Sect.~\ref{sec:paradigm}. However, such a weakening depends on the properties of the WD accretor, such as its mass and temperature, but is independent of the donor star transitioning from an M-dwarf to a brown dwarf. So, such systems would not necessarily experience an MB weakening at $M_2\approx 0.07 M_\odot$ but when the WD becomes magnetic (\citealt{2023A&A...679L...8S}).

\subsection{Dead zones}
The dead zone is the region around a spinning magnetised star in which the stellar wind is captured and forced to co-rotate along its magnetic field lines \citep{1987MNRAS.226...57M}. This leads to a reduction in wind mass loss and, as a consequence, the strength of MB. Dead zones were first studied by \cite{1987MNRAS.226...57M} who gave a simple description for isolated solar-like stars with different rotation rates. Subsequently, several groups have implemented the effects of dead zones in their calculations of MB torque in stellar spin-down \citep{2015ApJ...798..116R, Garraffo2015}.  Because dead zones arise through the interplay of gravity, centrifugal force, and magnetism in the star, they should be at play in every system undergoing MB. This includes the donor stars in CVs. The only difference here is that, owing to tidal locking, $P_\mathrm{orb}$ governs the behaviour of the dead zone. We calculated the evolution of dead zones using the simple treatment of \citet[their Eqs~(8) and (9)]{1987MNRAS.226...57M}, adopting solar parameters for the coronal temperature and mean molecular weight. {The choice of these parameters does not alter the qualitative behaviour of our dead zone calculations.} 

\begin{figure}
   \includegraphics[width=0.5\textwidth]{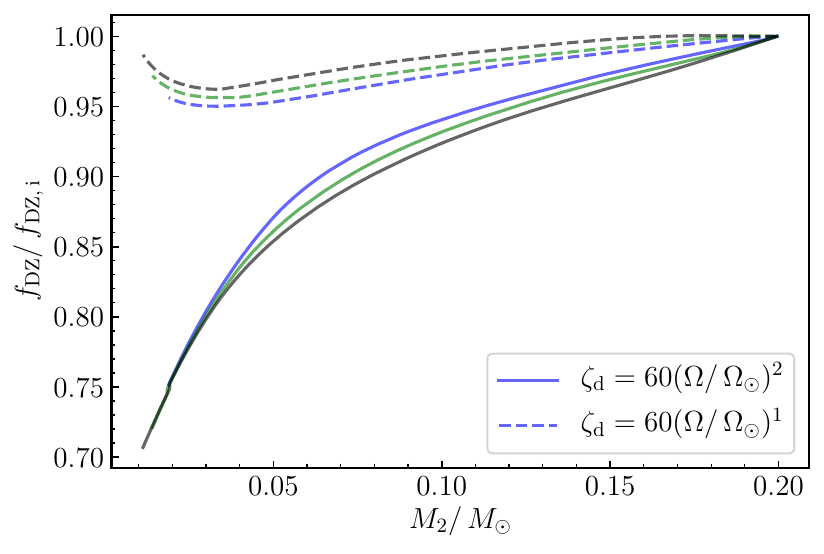}
   \caption{Evolution of the dead zone relative to that at the beginning of RLOF, $f_\mathrm{DZ}/f_\mathrm{DZ,\,i}$, with $M_2$ for the models with constant $\kappa$. {The colours denote the same models as in Figs~\ref{fig:P_M} and \ref{fig:p_mdot}, with $\kappa=0$ shown in blue, $\kappa=4$ in green, and $\kappa=15$ in black.} The line styles denote the choice of $\zeta_\mathrm{d}$. The dead zone for each track was calculated post-evolution using the method of \protect\cite{1987MNRAS.226...57M}.}
              \label{fig:dz}%
    \end{figure}

For the expression of the ratio of the magnetic pressure and the thermal pressure at the base of the dead zone, $\zeta_\mathrm{d}$, we can study the behaviour of two cases: $\zeta_\mathrm{d}=60(\Omega_/\Omega_\odot)$ and $\zeta_\mathrm{d}=60(\Omega/\Omega_\odot)^2$, from Table~1 of \cite{1987MNRAS.226...57M}. Here, $\Omega$ is the orbital angular velocity of the CV. The evolution of the dead zone of the donor star for the models with constant $\kappa$ in Fig.~\ref{fig:P_M} is shown in Fig.~\ref{fig:dz}. Here, $f_\mathrm{DZ}=R_2/R_\mathrm{DZ}$ is the fraction of field lines contributing to MB in the system, where $R_\mathrm{DZ}$ is the equatorial radius of the dead zone. With no dead zones, $f_\mathrm{DZ}=1$. The value $f_\mathrm{DZ,\,i}$ denotes the contribution of dead zones at the time of commencement of RLOF.\footnote{We note that this plot is made post-evolution so that dead zones do not alter the MB strength of these models.}  These tracks demonstrate how the dead zones would behave in a short-period CV. It is seen that when $\zeta_\mathrm{d}\propto \Omega$, $f_\mathrm{DZ}$ changes very little throughout the evolution. However, the dead zones grow ($f_\mathrm{DZ}$ becomes smaller) with decreasing $M_2$ when $\zeta_\mathrm{d}\propto \Omega^2$, with the drop becoming steep at $M_2\approx0.05M_\odot$. A stronger dependence of $\zeta_\mathrm{d}$ on $\Omega$ yields a steeper drop in $f_\mathrm{DZ}$. One way in which MB affects dead zones is through the generated magnetic field in the donor (say, by a strong-field dynamo or an $\alpha^2$ dynamo) that governs the magnetic pressure outside the star (through $\zeta_\mathrm{d}$). Dead zones work as an additional mechanism of MB alteration that is always at play regardless of the physical mechanism that drives MB. It can further weaken MB if $\mathrm{d\,ln}\,\zeta_\mathrm{d}/\,\mathrm{d\,ln}\,\Omega\gtrsim2$ (Fig.~\ref{fig:dz}).

\section{Conclusion}
\label{sec:conclusion}
In this Letter, we have shown that the weakening of MB in short-period CVs can \textcolor{black}{help} explain the dearth of observed period bouncers. \textcolor{black}{The main idea behind this stems from the fact that to reproduce the correct CV period minimum there ought to be some additional AML mechanism below the period gap. This need not necessarily be a constant scaling to GWR, as was employed by \cite{Knigge2011}. We introduce an MB at the lower end of the period gap that decreases as the CV approaches its period minimum and find that such a prescription also correctly reproduces the period minimum at about 80~min. In contrast to the systems with constant scaling, these systems spend considerable time around the observed period minimum spike between 80 and 86~min even after they have passed their minimum period. There, they become difficult to distinguish from pre-bounce systems. The mass-transfer rate decreases below the current detection threshold during further evolution driven by GWR and weakening MB. } 

\textcolor{black}{A direct comparison of our results with observations of the relative fraction of bouncers is difficult because of our ad hoc prescription of MB weakening and an uncertain detection threshold. So, we compare the constant MB models with the weakening MB ones for a range of detection thresholds and find that the latter shows a stronger reduction in the number of observable bouncers. Our models predict that the undetectable bouncers accumulate around $\dot{M}_2\approx10^{-12}M_\odot\,\mathrm{yr^{-1}}$, which can be tested with upcoming surveys that will hopefully reveal low-luminosity bouncers. }

The weakening of MB can be caused by physical processes such as a change in the dynamo action in the donor that drives weaker chromospheric activity, the emergence of magnetism in the white dwarf accretor that connects with that of the donor, and thus restricts the outflow of stellar winds, and dead zones in the donor trapping stellar winds. \textcolor{black}{Nevertheless, we admit that the paucity of observed period bouncers may be caused by other selection effects, too, or as-yet-unrecognised physical effects.} 

\begin{acknowledgements}
\textcolor{black}{We thank the referee for a detailed review which greatly improved the paper.} AS thanks the Gates Cambridge Trust for his scholarship. AS also thanks Ken Shen and Elmé Breedt for discussions on the nature of short-period CVs. ACR acknowledges support from an NSF Graduate Research Fellowship. AS and ACR are grateful to Franco Giovanelli and the Golden Age of Cataclysmic Variables and Related Objects VI Workshop for facilitating fruitful conversations. SG acknowledges support from the Israel Ministry of Innovation, Science, and Technology (grant No. 1001572596), and from the U.S. – Israel Binational Science Foundation (BSF; grant No. 2022175). AS and SG thank Daniel Blatman for the discussion on the emergence of magnetic fields in white dwarfs. CAT thanks Churchill College for his fellowship. 

\end{acknowledgements}

%
%

\bibliographystyle{aa} 
\bibliography{output}

\appendix

\section{Evolution of angular momentum loss}
\label{app:MB}

\begin{figure}
   \includegraphics[width=0.5\textwidth]{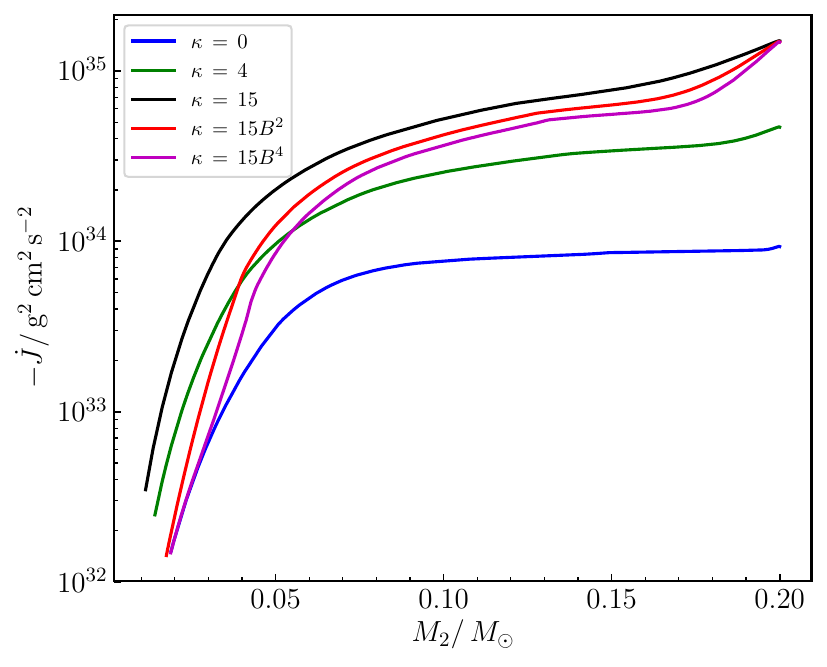}
   \caption{Evolution of $\Dot{J}$ with $M_2$ for the models in Fig.~\ref{fig:P_M}.}
              \label{fig:jdot}%
    \end{figure}

In Fig.~\ref{fig:jdot}, we plot the evolution of the total AML rate, $\Dot{J} \equiv \mathrm{AML_{GWR}\,+\, AML_{MB} } $, with the donor mass. We note that although $-\Dot{J}$ decreases throughout the evolution for each model, at $M_2\approx0.07M_\odot$ it decreases more steeply for the models in which $\kappa$ depends on stellar parameters.
    
\section{A catalogue of period bouncers}
\label{app:catalog}
We present a catalogue of all known period bouncers in Table \ref{tab:catalog}. While there are more such candidate systems in the literature (some 25 to 30 in total), we require there to be precise estimates of 1) an orbital period, 2) donor mass, 3) WD mass (and radius), and 4) WD temperature for us to include one in our sample. If a donor mass is not available, we ensure that WD properties are well measured and that there is spectroscopic evidence for a brown dwarf donor. We also indicate the few systems known to be eclipsing because those have, on the whole, more precisely measured donor star parameters.

\begin{table*}[]
    \caption{Known period bouncers with either a well-measured donor mass or accretion rate.}
\begin{tabular}{lrlllllll}
\toprule
\toprule
                    Name &  $P_\mathrm{orb}/$ hr & Ecl.? & $M_2/\,M_\odot$ & $M_\mathrm{WD}/\,M_\odot$ & $R_\mathrm{WD}/\,0.01\,R_\odot$ &$T_\mathrm{eff,\,WD}/\,$K & $\Dot{M}_2\,/10^{-10} M_\odot\,\textrm{yr}^{-1}$ &                               Ref. \\
\midrule
                  EZ Lyn &        1.430 &       Yes &        0.042$\pm$0.014 &  0.85$\pm$0.01 & 0.94 & 11250$\pm$40 &                     0.242$^{+0.003}_{-0.003}$ &                 1 \\
        SDSSJ1035 + 0551 &        1.370 &       Yes &        0.052$\pm$0.002 &  0.94$\pm$0.01 &  0.87$\pm$0.01 & 10100$\pm$200 &                      0.12$^{+0.005}_{-0.005}$ &                  2 \\
        SDSSJ1057 + 2759 &        1.510 &       Yes &       0.0436$\pm$0.002 &   0.80$\pm$0.015 &1.04$\pm$0.017 & 13300$\pm$1100 &                       0.54$^{+0.14}_{-0.17}$ &                  3 \\
        SDSSJ1433 + 1011 &        1.300 &       Yes &         0.06$\pm$0.003 &   0.868$\pm$0.007 & 0.958$\pm$0.008 & 12800$\pm$200 &                        0.38$^{+0.02}_{-0.02}$ &                  2 \\
SDSS J1501+5501 &        1.364 &       Yes &        0.053$\pm$0.003 &                0.80$\pm$0.03 & 1.04$\pm$0.04 & 12500$\pm$200 &         0.426$^{+0.009}_{-0.01}$ &                  2 \\
          SRGeJ0411+6853 &        1.625 &       Yes &            &           0.84$\pm$0.07  & 1.0$\pm$0.09 & 13790$\pm$500 &                0.56$^{+0.04}_{-0.04}$ &                   4 \\
                  EG Cnc &        1.439 &        No &             &      1.03$\pm$0.05 &0.77$^{+0.06}_{-0.05}$ & 12290$\pm$55 &                      0.2$^{+0.03}_{-0.03}$ &                        5 \\
                  GD 552 &        1.712 &        No &             &      0.78$\pm$0.04 &1.07$^{+0.05}_{-0.04}$ & 10760$\pm$40 &                     0.25$^{+0.02}_{-0.03}$ &                        5 \\
         1RXS J1050–1404 &        1.476 &        No &             &            0.77$\pm$0.03 &1.08$^{+0.04}_{-0.03}$ & 11520$\pm$50 &               0.34$^{+0.02}_{-0.03}$ &                        5 \\
                  QZ Lib &        1.539 &        No &             &      0.82$\pm$0.19 &1.01$^{+0.23}_{-0.18}$ & 11420$\pm$200 &                       0.28$^{+0.1}_{-0.2}$ &                        5 \\
         SDSS J1435+2336 &        1.300 &        No &             &           0.84$\pm$0.07 &1.0$^{+0.08}_{-0.09}$ & 12000$\pm$160 &                0.32$^{+0.05}_{-0.05}$ &                        5 \\
                  BW Scl &        1.304 &        No &        0.051$\pm$0.006 &     1.007$\pm$0.01 &0.8$^{+0.014}_{-0.011}$ & 15145$\pm$50 &                     0.51$^{+0.01}_{-0.008}$ & 5, 6 \\
                  GW Lib &        1.279 &        No &             &       0.83$\pm$0.12 &1.03$^{+0.15}_{-0.10}$ & 16166$\pm$350 &                    1.09$^{+0.26}_{-0.34}$ &                        5 \\
                  WZ Sge &        1.360 &        No &             &      0.80$\pm$0.02 &1.05$^{+0.03}_{-0.03}$ & 13190$\pm$115 &                     0.53$^{+0.01}_{-0.01}$ &                        5 \\

\bottomrule
\end{tabular}
\label{tab:catalog}
\textbf{Notes.} The accretion rates are derived from WD mass and temperature estimates (see Eq.~\ref{eq:townsley}).  Ecl. stands for eclipsing.\\ \textbf{References.} (1) \cite{2021amantayeva}; (2) \cite{2008littlefair}; (3) \cite{2017mcallister}; (4) \cite{2024arXiv240104178G}; (5) \cite{2022pala}; (6) \cite{2023neustroev}.
\end{table*}

\section{Computation of $\dot{M}_2$ and evolution in the $\dot{M}_2-P_\mathrm{orb}$ plane}
\label{app:accretion}
   \begin{figure*}
   \centering
   \includegraphics[width=1\textwidth]{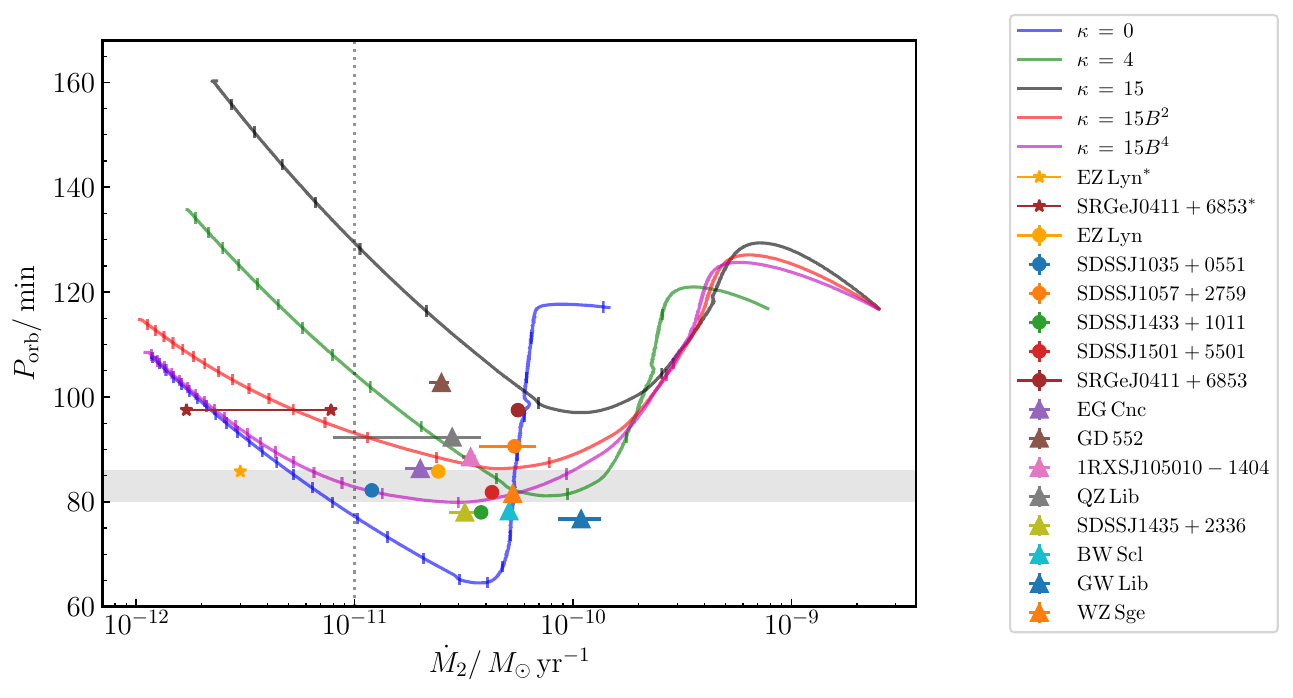}
   \caption{Evolution of CVs below the period gap. Solid lines show evolution in the $\Dot{M}_2-P_\mathrm{orb}$ plane for the same tracks as in Fig.~\ref{fig:P_M}. Ticks on each track denote timesteps of 300~Myr. The dotted vertical line denotes $\Dot{M}_2=10^{-11}M_\odot\,\mathrm{yr}^{-1}$. \textcolor{black}{During further evolution, the system is presumably undetectable (see text).} Observed period bouncer candidates from Table~\ref{tab:catalog} are also plotted. Eclipsing systems are plotted as circles, while non-eclipsing systems are plotted as triangles. {The systems labelled and marked with stars have their $\dot{M}_2$ derived from X-ray luminosity (Eq.~(\ref{eq:lum})), while the others have $\dot{M}_2$ derived from WD properties (Eq.~(\ref{eq:townsley})).} The horizontal shaded region is the observed period minimum spike ($80\lesssim P_\mathrm{orb}/\,\mathrm{min}\lesssim 86$) reported by \protect\cite{Gnsicke2009}. }
              \label{fig:p_mdot}%
    \end{figure*}
We computed $\Dot{M}_2$ using the relation derived given in \cite{2022pala},
\begin{gather}
    L_\textrm{WD} = 6\times10^{-3}L_\odot \left(\frac{\dot{M}_2}{10^{-10} M_\odot \textrm{ yr}^{-1}}\right)\left(\frac{M_\textrm{WD}}{0.9 M_\odot}\right)^{0.4}
    \label{eq:townsley}
\end{gather}
which relates the WD mass, radius, and temperature to the accretion rate (Table~\ref{tab:catalog}). It is also possible to estimate accretion rates from the X-ray luminosity or disc luminosity. The former requires a model of the X-ray emission mechanism, and the latter a model of disc geometry. Both require an estimate of accretion efficiency, which is often parametrised as $\eta$ in the following expression:
\begin{gather}
    L = \frac{\eta}{2} \frac{G M_\textrm{WD} \dot{M}_2}{R_\textrm{WD}}
    \label{eq:lum}
,\end{gather}
where $R_\mathrm{WD}$ is the WD radius and $L$ is the observed accretion luminosity (either from the disc or from the boundary layer, in X-rays). However, the range of $\eta$ in CVs is a subject of current debate (see Sect.~6.1 of \citealt{2017mukai} for a thorough explanation). As an example, one model of accretion is advective dominated accretion flow (ADAF), which was first applied to explain the hard X-ray spectra of CVs in \cite{1993narayan}. It was later extended in \cite{1996narayan} to X-ray binaries observed in a low accretion state. In this work, accretion efficiencies have been shown to be very low, with $\eta$ between $10^{-3}$ to $10^{-4}$. From Eq.~(\ref{eq:lum}), it is clear how failing to incorporate low efficiencies could lead to an underestimate of the accretion rate, given an observed luminosity. More recently, \cite{2008liu} applied the ADAF model to X-ray spectra of CVs and found good agreement. Nevertheless, \cite{2017mukai} warns that a complete analysis of accretion efficiency in CVs, which takes into account interactions between disc annuli, is still needed. 

\textcolor{black}{In Fig.~\ref{fig:p_mdot}, we describe how each model attains $\dot{M}_{2,\,\mathrm{detect}}$ in the $\Dot{M}_2-P_\mathrm{orb}$ plane. The model evolution can be explained as follows. Variation in the MB strength changes the $M_2-R_2$ relation, and consequently the $M_2-P_\mathrm{orb}$ relation of a CV, and so a CV attains $P_\mathrm{min}$ at a larger $P_\mathrm{orb}$ for a stronger MB (Fig.~\ref{fig:P_M}). Now, if MB stays constant post-bounce (say $\kappa=4$), then as CV bounces $P_\mathrm{orb}$ increases and $\dot{M}_2$ decreases. However, for the weakening MB model (say $\kappa=15B^4$), around the time when the CV bounces, MB becomes negligible. The CV now has to adjust its mass-transfer rate according to the current weak AML strength, but it cannot increase its $P_\mathrm{orb}$ because the $P_\mathrm{orb}$ where the CV bounced is too large for its AML strength. $\dot{M}_2$ has to reduce with little change in $P_\mathrm{orb}$.} So, the $\kappa=15B^4$ system essentially stays at approximately 80~min after bouncing before becoming undetectable (when $\Dot{M}_{2,\mathrm{detect}}=10^{-11} M_\odot\,\mathrm{yr}^{-1}$). The system stays in the period minimum spike (which is also populated by pre-bounce systems) before becoming undetectable. This is also seen in Fig.~\ref{fig:P_M} (the square in which $\dot{M}_{2,\,\mathrm{detect}}$ is attained is in the period minimum spike). This track explains observed candidates clustered at the lower end of the period minimum spike in Fig.~\ref{fig:p_mdot}. Similarly, the track with $\kappa=15B^2$ bounces at about 86~min but becomes undetectable at about 90~min. This track explains observed candidates clustered at the upper end of the period minimum spike. 
The system with $\kappa=4$ emerges from the period minimum spike with  $\Dot{M}_2>10^{-11}M_\odot \,\mathrm{yr}^{-1}$. So, if such a constant $\kappa$ is at play post-bounce, there should be systems populating the region with $86\lesssim P_\mathrm{orb}/\mathrm{min}\lesssim105$ and $\Dot{M}_2\gtrsim 10^{-11} \,M_\odot \,\mathrm{yr}^{-1}$. {These are not observed, indicating further that MB weakens post-period minimum.} If SRG/eROSITA unveils systems with $\Dot{M}_2\approx10^{-12}M_\odot \,\mathrm{yr}^{-1}$, the $\kappa=15B^4$ track indicates the existence of a population of systems up to $P_\mathrm{orb}\approx110\,$min and the $\kappa=15B^2$ track up to $P_\mathrm{orb}\approx115\,$min. \textcolor{black}{However, because $\eta$ in Eq.~(\ref{eq:lum}) is very uncertain and can easily be lower than even $10^{-4}$, such low-$\dot{M}_2$ systems would have very low luminosities. This can make finding them very difficult, even with the newer SRG/eROSITA surveys.}

{Finally, we note that our accretion rate estimates, based on WD properties (Eq.~(\ref{eq:townsley})), place the accretion rates of systems such as EZ~Lyn \citep{2021amantayeva} and SRGeJ0411+6853 \citep{2024arXiv240104178G} nearly an order of magnitude higher than that reported by authors using X-ray or disc luminosities. \cite{2021amantayeva} estimated the accretion rate based on the optical disc luminosity, and assumed that $\eta=1$ in Eq.~(\ref{eq:lum}) to obtain $\dot{M}_2\approx3\times 10^{-12}\,M_\odot\,\mathrm{yr^{-1}}$ (EZ~Lyn$^*$ in Fig.~\ref{fig:p_mdot}). \cite{2024arXiv240104178G} incorporated a bolometric correction to the X-ray luminosity, which assumed a thermal bremsstrahlung model for the emission, to obtain $\dot{M}_2\approx\,(1.7\,-\,7.8)\times 10^{-12}\,M_\odot\,\mathrm{yr^{-1}}$ (SRGeJ0411+6853$^*$ in Fig.~\ref{fig:p_mdot}). However, they did not explore a range of radiative efficiencies. In both cases, the accretion rates could have been underestimated. Another reason why these may have been underestimated is that $\dot{M}_2$ (lower end of SRGeJ0411+6853$^*$) is smaller than that for the $\kappa=0$ model. Assuming that the CV remains semi-detached, the estimates of the $\kappa=0$ model set the minimum accretion rate post-bounce. Regardless of these uncertainties, the $\dot{M}_2$ of EZ~Lyn$^*$ is only a factor of two smaller than that predicted by our $\kappa=15B^4$ model. It will agree with our model if we choose $\eta=0.5$ in Eq.~(\ref{eq:lum}) to calculate $\dot{M}_2$. The $\dot{M}_2$ of SRGeJ0411+6853$^*$ is already in general agreement with both the $\kappa=15B^2$ and $\kappa=15B^4$ models. 
Our model tracks agree well with several systems in Fig.~\ref{fig:p_mdot}, but notably the $\kappa=15B^4$ model is in good agreement with all the estimates of SDSSJ1501 and SDSSJ1035 — namely $P_\mathrm{orb}$, $M_2$, and $\Dot{M}_2$ — while the $\kappa=15B^2$ model is in agreement with the $P_\mathrm{orb}$ and $M_2$ estimate of EZ~Lyn and within a factor of two of its $\Dot{M}_2$ estimate.}

\end{document}